 \documentclass[letterpaper, 10 pt, conference]{ieeeconf}  

\usepackage{amsmath}
\usepackage{amsfonts}
\usepackage{amssymb,cite}
\usepackage{graphicx}
\usepackage{multirow}
\usepackage{makecell}
\usepackage{color}

\IEEEoverridecommandlockouts        

\title{\LARGE \bf
An Interpretable  Intensive Care Unit Mortality Risk Calculator
}

\author{Eugene T.~Y.~Ang$^{1}$,  Milashini Nambiar$^{2}$, Yong Sheng  Soh$^{1}$ and Vincent  Y.~F.~Tan$^{1}$
\thanks{$^{1}$Department of Mathematics, National University of Singapore
        {\tt\small e0175081@u.nus.edu, matsys@nus.edu.sg, vtan@nus.edu.sg}}%
\thanks{$^{2}$Institute for Infocomm Research, Agency for Science Technology and Research 
        {\tt\small Milashini\_Nambiar@i2r.a-star.edu.sg}}%
}

\begin{document}

\maketitle
\thispagestyle{empty}
\pagestyle{empty}

\begin{abstract}
Mortality risk is a major concern to patients have just been discharged from the intensive care unit (ICU). Many studies have been directed to construct machine learning models to predict such risk. Although these models are highly accurate, they are less amenable to interpretation and clinicians are typically unable to gain further insights into the patients' health conditions and the underlying factors that influence their mortality risk. In this paper, we use patients' profiles extracted from the MIMIC-III clinical database to construct  risk calculators based on different machine learning techniques such as logistic regression, decision trees, random forests and multilayer perceptrons. We perform an extensive benchmarking study that compares the most salient features as predicted by various methods. We observe a high degree of agreement across the considered machine learning methods; in particular, the cardiac surgery recovery unit, age, and blood urea nitrogen levels are commonly predicted to be the most salient features for determining patients' mortality risks. Our work has the potential for clinicians to interpret risk predictions. 

\end{abstract}

\section{Introduction}

Risk calculators are tools used by healthcare providers to estimate the probabilities of chronically ill or Intensive Care Unit (ICU) patients experiencing adverse clinical outcomes such as health complications, readmission to hospital, or mortality, and to then identify and screen high-risk patients.  In the context of patient mortality, severity scores such as SAPS-II \cite{SAPSII} and  sequential organ failure assessment (SOFA)~\cite{SOFA} have traditionally been used to predict hospital mortality.  These scoring systems predict mortality by feeding a fixed set of predictor variables, including various laboratory measurements and vitals, into simple models such as logistic regression.  The simplicity of these models limits their accuracy, and benchmarking studies \cite{benchmark} have found that automated risk calculators based on machine learning, that train supervised learning models on Electronic Health Record (EHR) data, tend to be far more accurate.

However, a drawback of using nonparametric machine learning models is that these models tend to be less interpretable than the simple models used in traditional scoring systems, i.e. the reasoning behind their predictions is not always transparent.  This paper investigates the problem of building a risk calculator that is not only accurate but interpretable.  We focus on the task of predicting the 30-day mortality of ICU patients at discharge, but note that the observations and principles applied to this work are broadly applicable to the tasks of predicting other adverse clinical outcomes using EHR data.

There is a considerable body of literature on the problem of ICU mortality prediction.  Typically, this literature uses data from the first 24 to 48 hours of a patient's stay to predict in-hospital mortality.  The 2012 PhysioNet Computing in Cardiology Challenge called for machine learning solutions to address the task of predicting in-hospital mortality for patients who stayed in the ICU for at least 48 hours \cite{physionetchallenge}.  The winning team developed a novel Bayesian ensemble learning solution and achieved an AUC of 0.86 \cite{winner}.  In the present paper, however, we study the task of 30-day mortality prediction at discharge---thus, the accuracies obtained are not directly comparable to the results in these other papers.  Another work that investigates risk prediction at discharge is \cite{dischargemortality}, in which the authors use LSTM models to predict the readmission of ICU patients within 30 days of their discharge, based on the last 48 hours of data from their stays.

A different stream of literature has looked at designing risk calculators that are not only accurate but interpretable.  A number of papers \cite{attention1,attention2} have proposed attention neural networks, where attention mechanisms are used to identify important features while retaining the accuracy of deep neural networks.  Some other works \cite{shapley1,shapley2,LIME} have also employed methods such as Shapley values, which are also used in this paper, and local interpretable model-agnostic explanation (LIME) to enhance interpretability.  More specifically, \cite{shapley1} combines Shapley values with XGBoost to predict the mortality of elderly ICU patients, while \cite{shapley2} combines Shapley values with Convolutional Neural Networks~\cite{cnn} to predict ICU mortality.   

A key difference between our paper and the literature on interpretable risk calculators is that rather than focusing on enhancing the interpretability of any particular machine learning method, we perform a benchmarking study that compares the reasoning of various methods.  We develop a series of risk calculators using Logistic Regression (LR), Decision Tree (DT), Random Forest (RF), $k$-Nearest Neighbors (k-NN), and Multilayer Perceptron (MLP); see Bishop's book \cite{bishop} for detailed descriptions of these methods.   In Section \ref{discussion}, we extract the most influential set of variables that each calculator relies on to make predictions. We observe that there is a high degree of commonality among these sets of variables, which suggests that all of these calculators rely on a similar set of  underlying reasonings to arrive at a prediction.  From a clinical perspective, our results are reassuring because they suggest that these data-driven methods for performing prediction in clinical settings make similar conclusions, even if the precise manner in which each method arrives at a decision can be quite different.

In Section \ref{sec:interpret}, we show how the set of influential factors derived in Section \ref{discussion} may be used to guide a clinician as to how a prediction was made. However, it is important for clinicians to understand the limitations of each model when using the risk calculator, so to gain better insights to the patients' health conditions. As LR constructs linear decision boundaries to distinguish patients' class, this model would not be that effective if the training data is severely not linearly separable. Even though the DT model is highly interpretable, slight changes in the training data may result in a completely different tree. Unlike DT, RF is more stable, however, it requires more computational resources to construct and aggregate various trees. Despite its simplicity, $k$-NN also requires a significant amount of memory to store all the training data and  to compute the prediction given a large dataset of patients' profiles. While the MLP can deal with training data that is not linearly separable, it is computationally intensive to train. It is  thus suggested for clinicians to use different models according to their needs.

\section{Methodology}

\subsection{Cohort Selection}

The data used in this study are extracted from the MIMIC-III Critical Care Database \cite{MIMIC}, which contains the health records of patients in Beth Israel Deaconess Medical Center ICU from 2001-2012.  For patients with multiple admissions, we considered every admission independently. There were a total of 61532 ICU records.

\subsection{Feature Extraction}

Our choice of features followed from studies done in the MIMIC-III research community and in MIT Critical Data~\cite{EHR}. In this study, our target variable was a binary flag, which indicated the patient's mortality within 28 days of their discharge from the ICU. For our input features, we selected 4 different data categories, namely, demographic, laboratory measurements, severity scores and vitals. 

For demographic features, we extracted the patients' height, weight, age, ethnicity, length of ICU stay, gender and the service unit that the patients were admitted to, namely the Coronary Care Unit (CCU), the Cardiac Surgery Care Unit (CSRU), the Medical Intensive Care Unit (MICU),  the Surgical Intensive Care Unit (SICU) and the Trauma/Surgical Intensive Care Unit (TSICU). We calculated the patients' BMIs by querying the patients' height and weight measurements that were last taken before the patients were discharged from the~ICU.

For laboratory measurement features, we extracted blood urea nitrogen (BUN), chloride, creatinine, hemoglobin, platelet, potassium, sodium, total carbon dioxide (TotalCO2) level and white blood cells (WBC) count of the patients. As these measurements might change due to the treatment the patients received during their ICU stay, we chose to query the patients' measurements that were last taken before the patients were discharged from the ICU.

We also calculated severity scores such as the patients' SOFA score \cite{SOFA} and the estimated Glomerular Filtration Rate (eGFR).  The latter was calculated based on the CKD-EPI Creatinine Equation~\cite{ckdepi}.

For vitals features, we extracted the temperature, heart rate, blood oxygen level (SpO2), systolic blood pressure (SysBP), diastolic blood pressure (DiasBP) and mean arterial pressure (MAP). Due to the non-stationary nature of the time series of vital signs, we chose to take the median value of the time series. 

\subsection{Data Processing}

Out of the 61532 records, the majority of the entries had missing height and weight features while a handful had missing features such as MAP, SpO2 and SysBP. Furthermore, there were anomalies in the certain features such as age and weight. We define an outlier as a value that is more than 3 standard deviations from the mean. We dropped records that contained these anomalies and those that had missing feature values, except for height, leaving 36330 data entries.

We then treated missing height entries as follows.  As is well known, the BMI can be expressed as the following function of the height and weight 
\begin{equation}\label{eqn 1}
    \mbox{BMI} = \mbox{weight}/\mbox{height}^{2}.
\end{equation}
We used ordinary least squares linear regression to regress BMI against weight and impute the missing height entries using the formula
\begin{equation}
    \widetilde{\mbox{BMI}} = 5.6925 + 0.2769 \times \mbox{weight}, \label{eqn:bmi}
\end{equation}
and the height features through Eqn.~\eqref{eqn 1}.    We found a strong positive correlation between BMI and weight, with an $R^{2}$  value of $0.737$, justifying that our imputation procedure is fairly accurate.  

An issue that we faced with the dataset was that the number of positively labelled records (patients who died within 28 days of discharge) represented 7.6\% of the total data sample, contributing to a severe class imbalance. One approach to tackle such class imbalance is Synthetic Minority Oversampling Technique (SMOTE) \cite{SMOTE}, which over-samples the examples in the minority class, by creating new examples from the minority class in the training data set.  This technique chooses a random minority class instance and finds its $k$ nearest minority class neighbours, based on the Euclidean distance. In our data processing step, we used the default setting for $k$, where $k = 5$, in the imblearn library. As the generated synthetic instance is a convex combination of that instance and one of its randomly chosen neighbours, it would likely contain some characteristics of the minority class due to its proximity (according to Euclidean distance).

As some of our features were categorical, we used SMOTE-NC (Synthetic Minority Over-sampling Technique for Nominal and Continuous) technique \cite{SMOTE} to deal with the categorical features in the dataset. SMOTE-NC sets the categorical values of the generated data by choosing the most frequent category of the nearest neighbours during the generation process. 

\subsection{Model Selection and Prediction}

After splitting our data so that 75\% of the original dataset is treated as the training data and the remaining 25\% as test data, we applied SMOTE-NC on the training set and trained the models $30$ times. 

The features were used to train LR, DT, RF, k-NN and MLP  in order to determine whether a patient would survive within 28 days of discharge from the ICU. The parameters of the model were estimated using 5-fold cross-validation (CV). The hyperparameters of each model were optimised through random and grid search. Also, in this optimisation, a different $5$-fold CV on the training set was performed. The tuned models were tested on the test set and performances were evaluated using the area under the receiving operating characteristic curve (AUC), test accuracy (ACC) and recall (REC). We then calculated the mean of the performance metrics across the 30 trials and used the standard deviation as the uncertainty bound. All modelling and analyses were performed with Python, and in particular the sklearn library.

\section{Model Performances} 

In Table \ref{modelperf}, we show the performance of our trained classifiers at predicting mortality of an ICU patient in the test data set.  With the exception of the Decision Tree model, all our trained classifiers attained a AUC $\geq 0.75$ or greater, and an accuracy (denoted by ACC) of $0.7$ or greater.  LR achieved the best performance in terms of AUC and REC, with $0.8$ and $0.713$ respectively, while the MLP prediction model had the best performance in terms of ACC at $0.8$. 
Our results suggest that all trained classifiers provide reasonable performance for predicting mortality.

\begin{table}[!h]
\begin{center}
\begin{tabular}{|c||c|c|c|c|c|}
\hline
& LR & DT & RF & kNN & MLP\\
\hline
AUC & \makecell{$\bold{0.8}$ \\ $\bold{\pm 0.008}$} & \makecell{0.696 \\ $\pm$ 0.02} & \makecell{0.764 \\ $\pm$ 0.007} & \makecell{0.761 \\ $\pm$ 0.008} & \makecell{0.76 \\ $\pm$ 0.009}\\
\hline
ACC & \makecell{0.744 \\ $\pm$ 0.004} & \makecell{0.675 \\ $\pm$ 0.065} & \makecell{0.742 \\ $\pm$ 0.007} & \makecell{0.713 \\ $\pm$ 0.005} & \makecell{$\bold{0.8}$ \\ $\bold{\pm 0.007}$}\\
\hline
REC & \makecell{$\bold{0.713}$ \\ $\bold{\pm 0.02}$} & \makecell{0.608 \\ $\pm$ 0.076} & \makecell{0.616 \\ $\pm$ 0.023} & \makecell{0.673 \\ $\pm$ 0.017} & \makecell{0.509 \\ $\pm$ 0.025}\\
\hline
\end{tabular}
\end{center}
\label{modelperf}
\caption{Performance on predicting ICU mortality on test data set across different classifiers.}
\end{table}

\section{Extracting Influential Features}\label{discussion}

Next, we extract the most influential features from each trained classifier.  We exhibit the high degree of overlap among the influential features across different classifiers.

\subsection{Logistic Regression}\label{logistic}

In LR, the trained classifier predicts mortality according to the following relationship
\begin{equation}\label{LogRegEqn}
    P(Y=1|\mathbf{x}) = 1/(1 + \exp\big(-(\boldsymbol{\theta}^T\mathbf{x}+\theta_{0}) \big)).
\end{equation}
Here, $\boldsymbol{\theta}$ represents the weights of the input features, $\mathbf{x}$ represents the patient's normalized data, and $\theta_{0}$ represents the offset of the decision boundary.  In particular, a larger value of $\boldsymbol \theta^{T} \mathbf{x} + \theta_0$ corresponds to a higher probability of mortality.  As such, we can infer the most influential features as those corresponding to the largest coefficients $\theta_i$ in magnitude.  In Table~\ref{LR-5}, we show the features corresponding to the five largest coefficients and the five smallest coefficients.  

\begin{table}[!h]
\begin{center}
\begin{tabular}{|c||c|c|c|c|}
\hline
\multicolumn{5}{|c|}{Results}\\
\hline
& \multicolumn{2}{|c|}{Positive Influence} & \multicolumn{2}{|c|}{Negative Influence}\\
\hline
1 & Age & 0.586 & CSRU & -1.176\\
\hline
2 & BUN & 0.336 & TSICU & -0.444\\
\hline
3 & Male & 0.325 & SICU & -0.296\\
\hline
4 & MICU & 0.285 & BMI & -0.288\\
\hline
5 & Heart Rate & 0.273 & Hemoglobin & -0.198\\
\hline
\end{tabular}
\end{center}
\caption{Top 5 most influential features by logistic regression}\label{LR-5}
\end{table}

We observed from Table \ref{LR-5} that all service units except MICU had a negative influence on the mortality risk. We also observed that older patients, male patients, patients with higher blood urea nitrogen (BUN) or heart rate would tend to have a greater mortality risk within 28 days after their discharge. These features are possible indicators for poor renal or cardiovascular functions. Similarly, patients that had higher BMI or hemoglobin level would tend to have lower mortality risk.

To increase the interpretability of the model, we penalize the logistic regression problem with an $\ell_1$-norm regularizer, i.e., we solve
\begin{equation}
    \min_{\boldsymbol{\theta},\theta_{0}} \sum_{t=1}^{n}\log \Big(1+\exp\big(-y_{t}(\boldsymbol{\theta}^T\mathbf{x}_{t}+\theta_{0})\big)\Big) + \lambda \|\boldsymbol{\theta}\|_{1}.
\end{equation}
Here, $\lambda>0$ is the regularization parameter, which we tune via CV. 

After tuning the hyperparameters with 5-fold CV and running the model $30$ times, the best estimator achieved a test score of $0.72  \pm  0.005$, recall of $0.734  \pm  0.021$ and AUC of $0.797  \pm  0.009$.  We annihilated those features which had a coefficient of zero more than half of the time \cite{stats}.

\begin{table}[!h]
\caption{FEATURES WITH NON-ZERO COEFFICIENTS IN L1-PENALISED LR}
\begin{center}
\begin{tabular}{|c||c|c|}
\hline
\multicolumn{3}{|c|}{Results}\\
\hline
 & Features & Coefficients \\
\hline
1 & Age & 0.494 $\pm$ 0.017\\
\hline
2 & BUN & 0.236 $\pm$ 0.014\\
\hline
3 & SOFA & 0.236 $\pm$ 0.014\\
\hline
4 & Heart Rate & 0.22 $\pm$ 0.017\\
\hline
5 & MICU & 0.101 $\pm$ 0.059\\
\hline
6 & WBC & 0.071 $\pm$ 0.016\\
\hline
7 & Length of stay & 0.063 $\pm$ 0.017\\
\hline
8 & Male & 0.023 $\pm$ 0.035\\
\hline
9 & MAP & -0.011 $\pm$ 0.012\\
\hline
10 & SysBP & -0.016 $\pm$ 0.014\\
\hline
11 & Temperature & -0.1 $\pm$ 0.015\\
\hline
12 & Hemoglobin & -0.123 $\pm$ 0.019\\
\hline
13 & BMI & -0.198 $\pm$ 0.019\\
\hline
14 & CSRU & -1.362 $\pm$ 0.063\\
\hline
\end{tabular}
\end{center}
\label{annihilate}
\end{table}

We observed that $11$ features were annihilated. Thus, clinicians could determine to mortality risk by the sparse selection of influential features as shown in Table~\ref{annihilate}.

\subsection{Decision Tree}\label{decision}

To obtain an interpretable model, we restricted the learned DT model to a maximum depth of 5 tiers. One approach to find out which features are more influential in a DT model is to compute the Gini Importance of each feature.  This is given as the decrease in Gini impurity weighted by the node probability where the {\em Gini impurity} of a certain node is
\begin{equation}
    G(p_1, p_2) = 1 - p_{1}^{2}-p_2^2,
\end{equation}
where 
$p_i$ is the probability of picking a data instance from class $i \in\{1,2\}$ in that node. Hence, the larger the feature's Gini importance, the more important the feature is \cite{Gini}. The top 5 features with the highest Gini importances are shown in Table~\ref{Gini5}. 
\begin{table}[!h]
\caption{TOP 5 FEATURES OF THE HIGHEST GINI IMPORTANCE}
\begin{center}
\begin{tabular}{|c||c|c|}
\hline
\multicolumn{3}{|c|}{Results}\\
\hline
 & Features & Gini Importance \\
\hline
1 & Length of stay & 0.224773\\
\hline
2 & CSRU & 0.214003\\
\hline
3 & BUN & 0.174614\\
\hline
4 & Age & 0.052457\\
\hline
5 & eGFR & 0.047148\\
\hline
\end{tabular}
\end{center}
\label{Gini5}
\end{table}

We observed from Table \ref{Gini5} that length of stay, service unit CSRU, blood urea nitrogen level, age and estimated GFR had the highest Gini Importance. This means that most of the decisions that were made on how to split the data were mostly based on these features, and that the samples in the following nodes were relatively pure. Hence, clinicians could obtain a reasonable prediction of the risk of mortality by looking just at this smaller subset of features.

\subsection{Random Forest}\label{forest}
A random forest is an ensemble of decision tree models, and is hence less amenable to interpretation. As such, we compute the Shapley value of each feature to infer its influence.  The concept of Shapley values were first introduced in the field of cooperative game theory, and it measures the marginal contribution that each feature (player, in the context of game theory) to a subset of features (coalition), averaged across all possible subsets of features \cite{Game}. The Shapley value of $i$-th feature is given as
\begin{equation}
    \phi_{i}( {f}) \!=\!\! \sum_{ S \subseteq \{x_{j} \}_{j=1}^p
    \setminus\{x_{i}\}} \!\!\!\!\frac{|S|!(p\!-\!|S|\!-\! 1)!}{p!}( {f}(S \cup \{x_{i}\})- f (S)).
\end{equation}
Thus $\phi_{i}(f)$ is the contribution of the $i$-th feature based on ${f}$, which calculated the economic output of all feature values in $S$, a subset of the features used in the model, $p$ is the number of features and $x_{i}$ represents the corresponding feature values of a data point to be explained. 
Intuitively, a feature has a higher Shapley value if the inclusion of this feature in the prediction model results in a greater change in predictions, as compared to the inclusion of other features.

For every data instance, the average marginal contribution of each feature is given by its Shapley value. In order to find out the influence of the feature, we can determine the SHAP feature importance by averaging the Shapley value for every features across all data instances \cite{shapley}.

\begin{figure}[h]
	\centering
	\includegraphics[width=8cm]{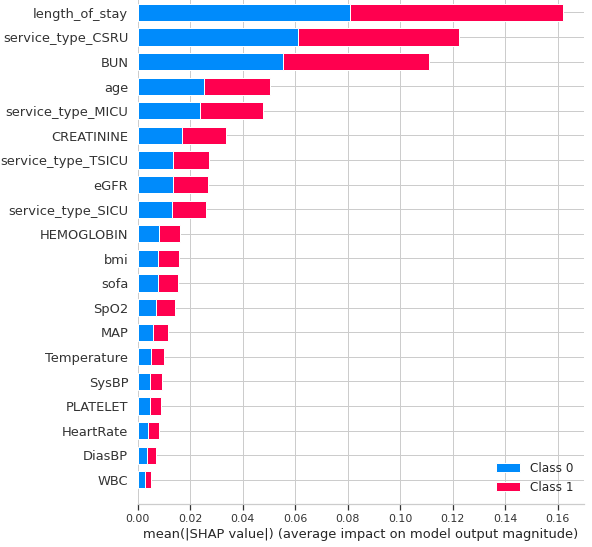}
	\caption{SHAP feature importance for Random Forest}
	\label{shaprf}
\end{figure}

Note that the Shapley value requires us to sum over all possible subsets of features.  This becomes intractable if the number of features is moderately large. In our numerical implementations, we approximately compute the Shapley value by averaging over a small number of randomly sampled subsets of features. For the RF model, we implemented a TreeExplainer from the shap library, which the class disregarded decision paths that involved missing features instead of computing the output through a selection of the features~\cite{treeexplainer}. In Fig.~\ref{shaprf} we show the SHAP feature importance values across the top 20 features.

We observed from Fig.~\ref{shaprf} that length of stay had the largest SHAP feature importance and was responsible for changing the predicted absolute ICU mortality risk prediction on average by around 8\% (0.08 on x-axis) regardless of classes. As the larger the SHAP feature importance, the bigger the average marginal contribution to risk prediction the feature had. We could also observe that length of stay, service unit CSRU, blood urea nitrogen level, age and service unit MICU were the 5 most influential features. In fact, for many instances, clinicians could accurately deduce the final predicted risk mortality with our trained RF model with these 5 features while choosing the remaining features randomly.

\subsection{k-Nearest Neighbor}\label{knn}

As our implementation of $k$-NN uses weights trained using the RF model, the most influential variables using $k$-NNs are precisely those found using these other methods.  As such, we omit discussing $k$-NNs.



\subsection{Multilayer Perceptron}\label{perceptron}

\begin{figure}[h]
	\centering
	\includegraphics[width=8cm]{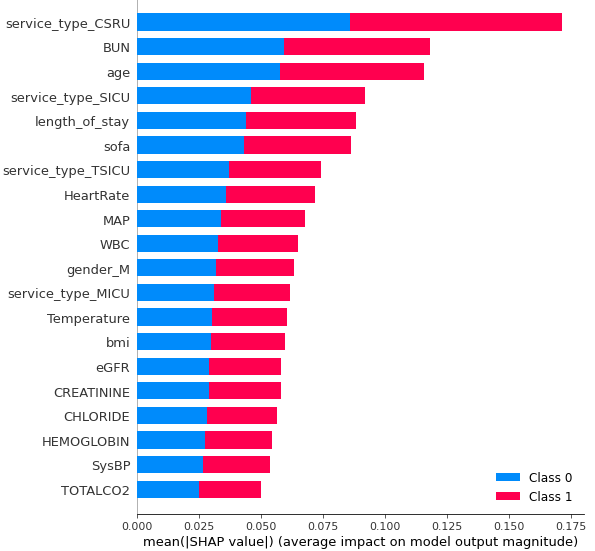}
	\caption{SHAP feature importance for MLP}
	\label{shapmlp}
\end{figure}

The MLP is a black-box model, and hence is less amenable to interpretation.  Hence, we computed Shapley value using the KernelExplainer from the shap library to interpret the model \cite{kernel}. In order to compute the SHAP feature importance, we had to take the average of the Shapley values of every features across all data instances, so to reduce the computational complexity, we estimated the SHAP feature importance from 500 training data instances randomly and repeated the process 30 times. We plot the SHAP feature importance values for MLP in Fig. \ref{shapmlp}.

We observed from Fig. \ref{shapmlp} that service unit, CSRU had the largest SHAP feature importance and was responsible for changing the predicted absolute ICU mortality risk prediction on average by around 8\% regardless of classes. Similar to the analysis of the RF model, clinicians could accurately deduce the final predicted mortality risk by looking at the 5 most influential features, namely service unit CSRU, blood urea nitrogen level, age, service unit SICU and length of stay while choosing the remaining features randomly.

\subsection{Comparison of Influential Features}

We compare the top few most influential features extracted from the various models, and we tabulate these in Table \ref{4models}.  We observed a high degree of agreement in the most important features across all models; in particular, we observed that service unit CSRU, age, and blood urea nitrogen levels ranked among the top 5 most influential features across all models.  This suggested that these models made predictions with largely similar reasonings.


\begin{center}
\begin{table}[h]
\begin{center}
\begin{tabular}{|c|c|c|c|c|}
\hline
& LR & DT & RF & MLP\\ 
\hline 
$1$ & CSRU & Length of stay & Length of stay & CSRU\\
\hline
$2$ & Age & CSRU & CSRU & BUN\\
\hline
$3$ & TSICU & BUN & BUN & Age\\
\hline
$4$ & BUN & Age & Age & SICU\\
\hline
$5$ & Gender (M) & eGFR & Creatinine & Length of stay\\
\hline
\end{tabular}
\end{center}
\caption{Top 5 ranked influential features across different models}
\label{4models}
\end{table} 
\end{center}

\section{Interpreting Risk Predictions}\label{sec:interpret}

Next, we apply the influential factors obtained in Section \ref{discussion} to interpret the predictions derived from our risk calculators. We focus on 4 specific test cases, Patients \#5, \#17, \#53, and \#154, from the test data set. Patients \#53 and \#154 did not survive past 28 days after discharge, while Patients \#5 and \#17 survived past 28 days after discharge.

\textbf{Logistic Regression.}  We recall that LR predicts the mortality risk of a patient based on the linear function $\boldsymbol{\theta}^T\mathbf{x}+\theta_{0}$.  Hence the final prediction is primarily influenced by how far these feature values deviate from the population sample, with its importance weighted by the coefficients in  $(\boldsymbol{\theta},\theta_0)$.  As such, a clinician who wishes to understand whether the LR's prediction is sound can examine how far the patient's feature values deviate from the population, with special attention on features whose coefficients have larger maginitude.  

As an illustration, we show the numerical values of each patient corresponding to the influential features, as listed in Table \ref{annihilate} in our fitted LR model in Table \ref{4examplesLR}.
We observed that Patient \#53 had a high heart rate and a low systolic blood pressure, mean arterial pressure and hemoglobin level, relative to the population. These features contributed towards predicting a high mortality risk, and in this instance, the patient did not survive.
We observed that Patient \#17 had a relatively high BUN, SOFA score, systolic blood pressure and BMI.  The elevated BUN and SOFA score, which have positive coefficients, outweighed the elevated systolic pressure and BMI numbers, which have negative coefficients.  As a result, our model predicted an overall elevated mortality risk, even though the patient did survive past the 28 days.  

\begin{center}\label{4examplesLR}
\begin{table}[h]
\caption{RAW FEATURES VALUES OF THE 4 EXAMPLES}
\begin{center}
\resizebox{\linewidth}{!}{
\begin{tabular}{|c|c|c|c|c|c|}
\hline
\multirow{2}{*}{Features}&
\multirow{2}{*}{Range}&
\multicolumn{4}{c|}{Test Examples} \\ \cline{3-6}& & $\#53$ & $\#154$ & $\#17$ & $\#5$\\ 
\hline
\multicolumn{2}{|c|}{Service Unit} & CCU & SICU & MICU & CSRU\\
\hline
Age (years) & $63.2 \pm 16.2$ & $73.7$ & $52.2$ & $70.53$ & $57.04$\\
\hline
BUN (mg/dL) & $22.4 \pm 15.6$ & $28$ & $19$ & $\bold{54}$ & $13$ \\
\hline
\multicolumn{2}{|c|}{Male} & $0$ & $1$ & $0$ & $0$ \\
\hline
Heart Rate (bpm) & $83.9 \pm 13.8$ & $\bold{108}$ & $82$ & $82$ & $81.5$\\
\hline
SOFA & $4.1 \pm 3$ & $7$ & $2$ & $\bold{8}$ & $5$\\
\hline
WBC (K/$\mu$L) & $10.5 \pm 4.7$ & $8.3$ & $11.7$ & $13.1$ & $13.7$\\
\hline
Length of stay (days) & $3.9 \pm 6.4$ & $3$ & $1$ & $7$ & $1$\\
\hline
MAP (mmHg) & $77.9 \pm 10.2$ & $\bold{62}$ & $86.5$ & $73$ & $75.5$\\
\hline
SysBP (mmHg) & $119.9 \pm 15.9$ & $\bold{100}$ & $132.5$ & $\bold{138}$ & $106$\\
\hline
Temperature (Celsius) & $36.9 \pm 0.5$ & $36.4$ & $\bold{37.8}$ & $37.1$ & $36.8$\\
\hline
Hemoglobin (g/dL) & $10.5 \pm 1.7$ & $\bold{8.3}$ & $12$ & $10$ & $9.2$\\
\hline
BMI (kg m$^{-2}$) & $28.1 \pm 6.4$ & $32.7$ & $27.6$ & $\bold{38.5}$ & $\bold{43.49}$\\
\hline
\end{tabular}}
\end{center}
\end{table} 
\end{center}

\textbf{Decision Tree.}  A DT model makes a prediction by answering a sequence of queries, which checks if a particular feature value is above or below a learned threshold.  Hence, a clinician who wishes to understand if the learned DT model makes decisions in a fashion that is clinically sound can study these queries, and in particular, check if these queries can be backed by clinical expertise.
The respective branches of all examples are shown in Table \ref{4examplesDT}. 

We observed from Table \ref{4examplesDT} that Patients \#53 and \#17 undertook the same sequence of branches under our learned model even though they had different outcomes.  One possible reason for the incorrect classification may be that the learned DT model has limited classification power because it only has 5 levels.  Another reason may be that DT models are inherently limited in that they on hard decision boundaries, and  cannot distinguish between features that are near or far away from a decision boundary.  In some cases, the severity of a certain feature value may sometimes convey clinically relevant information.

\begin{center}
\begin{table}[ht]
\begin{center}
\resizebox{\linewidth}{!}{
\begin{tabular}{|c|c|c|c|c|}
\hline
Nodes & $\#53$ & $\#154$ & $\#17$ & $\#5$\\
\hline 
$1$ & BUN $> 19$ & BUN $\leq 19$ & BUN $> 19$ & BUN $\leq 19$\\
\hline
$2$ & TSICU $\leq 0.5$ &  Length of stay $\leq 2$ & TSICU $\leq 0.5$ & Length of stay $\leq 2$\\
\hline
$3$ & CSRU $\leq 0.5$ & CSRU $\leq 0.5$ & CSRU $\leq 0.5$ & CSRU $> 0.5$\\
\hline
$4$ & SpO2 $\leq 99.9$ & Age $\leq 52.6$ & SpO2 $\leq 99.9$ & -\\
\hline
$5$ & Hemoglobin $\leq 11.9$ & Age $> 43.7$ & Hemoglobin $\leq 11.9$ & -\\
\hline
\end{tabular}}
\end{center}
\caption{DT branches of Patients \#5, \#17, \#53, and \#154}
\label{4examplesDT}
\end{table} 
\end{center}

\begin{figure}[h]
	\centering
	\includegraphics[width=8cm]{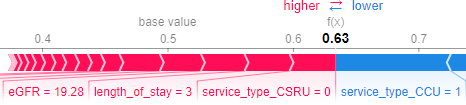}
	\caption{SHAP explanation force plot for Patient \#53}
	\label{shaprf53}
\end{figure}

\begin{figure}[h]
	\centering
	\includegraphics[width=8cm]{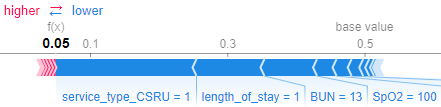}
	\caption{SHAP explanation force plot for Patient \#5}
	\label{shaprf5}
\end{figure}

\textbf{Random Forest.}  As we noted in Section \ref{forest}, RF is an ensemble method, and hence its results are less amenable to direction interpretation compared to LR and DT.  As such, we rely on the SHAP explanation forces to interpret the predictions provided by the RF model.  Broadly speaking, Shapley values can be viewed as ``forces'', which either increase or decrease the mortality risk. Each Shapley value is represented as an arrow that pushes to increase or decrease the prediction probability and the magnitude of the contribution is showed by the size of the arrow. These forces would balance each other out at the actual prediction of the data instance. The SHAP explanation force plots for the examples \#53 and \#5 are shown in Figs. \ref{shaprf53}, \ref{shaprf5} respectively.

We observed from Figs.~\ref{shaprf53} and~\ref{shaprf5} that the baseline probability, which was the average predicted probability, was $0.5$. We observe from Fig.~\ref{shaprf53} that patient \#53 had an elevated predicted risk of $0.63$. Also, the marginal risk-decreasing contribution of being in the service unit CCU is the largest, according to the size of the arrow. Although the 3 most influential risk-increasing factors such as not being in service unit, CSRU, 3 days in the ICU and an estimated GFR of 19.28 mL/min/1.73m$^{2}$ has less absolute marginal contribution than the service unit CSRU, the combined effects of these factors outweighed the risk-decreasing factor and thus, leads to an overall higher risk prediction.





We   observe from Fig.~\ref{shaprf5} that patient \#5 had a low predicted risk of $0.05$. Also, the marginal risk-decreasing contribution of being in the service unit CSRU is the largest, according to the size of the arrow. The risk-increasing factors had negligible marginal  contribution to the overall risk prediction, whereas risk-decreasing factors such as being in the service unit CSRU, 1 day in the ICU, blood urea nitrogen level of 12 mg/dL have large absolute marginal contribution, as observed from the size of the arrows. This leads to an exceptionally low overall mortality risk. 


\textbf{$k$-Nearest Neighbors.}  The $k$-Nearest Neighbors model makes prediction based on the outcomes of the nearest neighbors. As the best estimator used 16 neighbors to make a prediction, we showed the feature values of the patient \#53 and the range of its $16$ neighbors in Table \ref{knn53154}. 

We observed from Table \ref{knn53154} the majority of the neighbors had a positive label,   resulting in a positive prediction for test example \#53. Also, we observed that the feature values of the test sample were generally close to the range of its neighbours despite a few anomalies. This was likely due to a large range in the raw values of certain features such as platelet count. This resulted in a large disparity even though the standardised values were similar.

\begin{center}
\begin{table}[h]
\begin{center}
\resizebox{\linewidth}{!}{
\begin{tabular}{|c||c|c|}
\hline
Features & $\#53$ & Range of 16 neighbors' feature values\\
\hline 
Age & 73.7 & $73.7 \pm 5.67$ \\
\hline
Length of stay & 3 & $3.96 \pm 1.39$  \\
\hline
SOFA & 7 & $5.79 \pm 1.68$  \\
\hline
DiasBP & 51 & $53.5 \pm 3.38$ \\
\hline
HeartRate & 108 & $99.3 \pm 5.89$  \\
\hline
MAP & 62 & $65.9 \pm 3.41$ \\
\hline
SpO2 & 96 & $96.8 \pm 0.84$\\
\hline
SysBP & 100 & $96.3 \pm 6.58$ \\
\hline
Temperature & 36.4 & $36.7 \pm 0.18$ \\
\hline
BUN & 28 & $37.4 \pm 13.9$\\
\hline
Chloride & 92 & $94.5 \pm 2.6$ \\
\hline
Creatinine & 2.4 & $2.22 \pm 0.447$ \\
\hline
Hemoglobin & 8.3 & $9.37 \pm 0.459$ \\
\hline
Platelet & 266 & $223 \pm 121$ \\
\hline
Potassium & 4 & $4.27 \pm 0.42$ \\
\hline
Sodium & 133 & $134 \pm 2$ \\
\hline
TotalCO2 & 31 & $31.1 \pm 1.28$ \\
\hline
WBC & 8.3 & $8.41 \pm 1.66$ \\
\hline
BMI & 32.7 & $29.8 \pm 3.8$ \\
\hline
eGFR & 19.3 & $29.8 \pm 7.62$\\
\hline
Gender & F & 14 male, 2 female\\
\hline
Service Unit & CCU & 14 CCU, 2 MICU\\
\hline
Label & Positive & 15 positive, 1 negative\\
\hline
\end{tabular}}
\end{center}
\caption{Feature values of example \#53 and its 16 neighbors}
\label{knn53154}
\end{table} 
\end{center}

\begin{figure}[h]
	\centering
	\includegraphics[width=8cm]{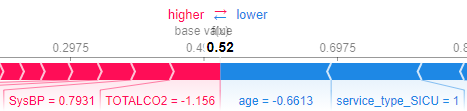}
	\caption{SHAP explanation force plot for Patient \#154}
	\label{shapmlp154}
\end{figure}

\textbf{Multilayer Perceptron.}  Our analysis for MLP is based on examining SHAP explanation force plots, and is similar to our discussion regarding the RF model. The SHAP explanation force plots for the example \#154 is shown in Figure~\ref{shapmlp154}.



We observed from Fig.~\ref{shapmlp154} that the base value, which was the average predicted probability, was $0.5$. We observe from Fig.~\ref{shapmlp154} that patient \#154 had a predicted risk of $0.52$. Also, the marginal risk-decreasing contribution of age, with a raw value of $52.2$ is the largest, according to the size of the arrow. Although risk-decreasing factors such as age and being in the service unit SICU have large marginal contribution, these contributions are counteracted by several less influential risk-increasing factors such as the systolic blood pressure and total carbon dioxide level. Hence, the MLP model may not be able to provide a definitive prediction on this patient's mortality risk and clinicians could consider using other models such as DTs or k-NNs to estimate the mortality risk.





\section{Conclusion}

In this study,  based on patients' profiles extracted from  the MIMIC-III clinical database,  we   developed various risk calculators to predict 30-days mortality risk of ICU patients at discharge. Not only have we have not only determined the common influential features that all calculators relied on to make predictions, but we have also developed various methods to interpret the risk predictions. On top of achieving high performance on the test data, these calculators shared a high degree of commonality among the set of influential features. This showed the effectiveness of these risk calculators in clinical settings. As this study can be helpful in providing interpretable yet accurate predictions for clinicians, future research can be done on exploring {\em causal} models. This has the potential to provide clinicians with insights regarding the causal relationships between these features and thus improving the patients' health conditions by targeting the independent variables.

\addtolength{\textheight}{-12cm}   



%




\end{document}